\newcommand{\apj}{\textit{ApJ}}
\newcommand{\apjl}{\textit{ApJ}}
\newcommand{\aap}{\textit{A\&A}}
\newcommand{\solphys}{\textit{Sol. Phys.}}

\newcommand{\jgr}{{\textit J. Geophys. Res.}}

\documentclass{iau}
\usepackage{graphicx}

\title[Successive filament eruptions within the breakout model] 
{Successive filament eruptions within one solar breakout event}

\author[Y. Shen]   
{Yuandeng Shen$^{1,2}$}

\affiliation{$^1$Yunnan Astronomical Observatory, Chinese Academy of Sciences, Kunming 650011, China\\
$^2$Kwasan and Hida Observatories, Kyoto University, Kyoto 6078471, Japan \\ email: {\tt ydshen@ynao.ac.cn} \\[\affilskip]}

\pubyear{2013}
\volume{8 Symposium S300}  
\pagerange{231--234}
\jname{IAUS300: Nature of prominences and their role in space weather}
\editors{A.C. Editor, B.D. Editor \& C.E. Editor, eds.}
\begin{document}

\maketitle

\begin{abstract}
The magnetic breakout model has been widely used to explain solar eruptive activities. Here, we apply it to explain successive filament eruptions occurred in a quadrupolar magnetic source region. Based on the  high temporal and spatial resolution, multi-wavelengths observations taken by the Atmospheric Imaging Assembly (AIA) on board the {\it Solar Dynamic Observatory} ({\it SDO}), we find some signatures that support the occurrence of breakout-like external reconnection just before the start of the successive filament eruptions. Furthermore, the extrapolated three-dimensional coronal field also reveals that the magnetic topology above the quadrupolar source region resembles that of the breakout model. We propose a possible mechanism within the framework of the breakout model to interpret the successive filament eruptions, in which the so-called magnetic implosion mechanism is firstly introduced to be the physical linkage of successive filament eruptions. We conclude that the structural properties of coronal fields are important for producing successive filament eruptions.
\end{abstract}

\firstsection 
\section{Introduction}
Filament (prominence) eruption is one of the most spectacular, large-scale activity on the Sun, which often associates with solar flare and coronal mass ejection (CME). The eruption of a filament can severely impact the solar-terrestrial environment and human activities; and the study of these phenomena has developed into a new discipline dubbed space weather. However, the physical mechanism of filament eruption is still not well understood, even though extensive observational and theoretical works have been made in recent decades.

Generally speaking, a filament eruption always starts from a closed magnetic system in quasi-static equilibrium, in which the upward magnetic pressure force of the low-lying sheared field is balanced by the downward tension force of the overlying field. When the eruption begins, the equilibrium is destroyed catastrophically, and part of the non-potential magnetic flux and the plasma are expelled violently from the Sun. Given different magnetic environments, the eruption of a filament can be failed, partial, and complete eruptions (e.g., \cite{gilb01,liu09,shen11}). Sometimes, several filaments far from each other or resided in one complex active region can erupt successively within a short time period. A key question of successive eruptions is whether they are physically connected or not. It seems that the answer is positive, and the connection is often of a magnetic nature (e.g., \cite{jian08,jian11, toro11,shen12,tito12,lync13,schr11,schr13}).

Currently, solar physicists have developed a number of models for interpreting filament/CME eruptions. Among various models, the magnetic breakout model assumes a large-scale quadrupolar field configuration; the core field is increasingly sheared by photospheric motions, which is surrounded by an overlying antiparallel loop system (\cite{anti98,anti99}). Naturally, a null point is formed between the core and the overlying loop system. This model can be used to interpret many eruptions which occur in complex multipolar active regions (e.g., \cite{aula00,maia03,shen12}). Here, we apply the magnetic breakout model to explain successive filament eruptions which occurred in a quadrupolar magnetic source region, and propose a possible physical linkage between the filament eruptions.

\section{Results \& Interpretation}
\begin{figure}[b]
\begin{center}
\includegraphics[width=0.8\textwidth]{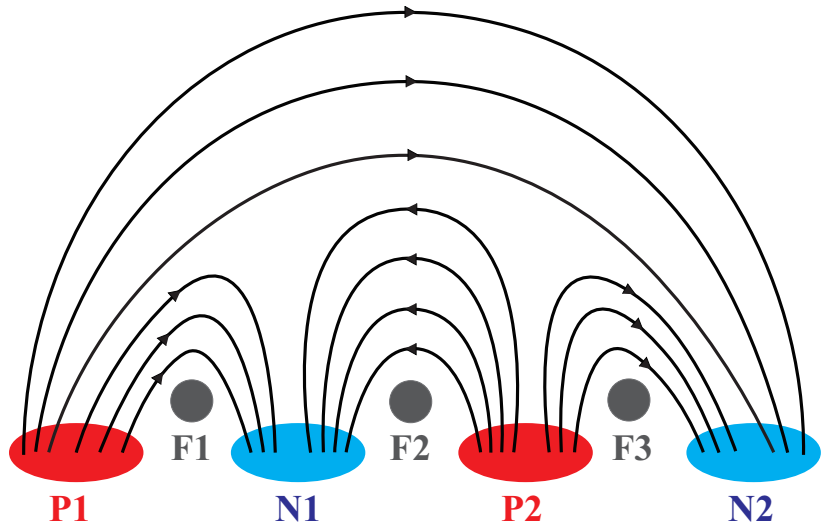} 
\caption{Topology structure of the breakout mode. The red (P1 and P2) and blue (N1 and N2) patches indicate the positive and negative polarities of the magnetic field, respectively. The field direction is indicated by a series of arrows. The gray patches represent the filaments (F1, F2, and F3) confined by the three low-lying lobes.}
\label{fig1}
\end{center}
\end{figure}

The basic magnetic topology of the breakout model is shown in Figure.\,\ref{fig1}. It can be seen that the four magnetic poles (P1, N1, P2, and N2) are connected by three low-lying lobes and one overlying loop system; and a coronal null resides inbetween the middle lobe and the overlying antiparallel loop system. In addition, one can assume the existence of a filament under each lobe (F1, F2, and F3). In such a configuration, a small disturbance to the system could lead to the eruption of the whole system. Typically, there are two types of disturbance to the system. The first type is that the disturbance acts on F2, which will lead to the rising of F2 and the middle lobe, and further results in the external reconnection around the null point, which will removes the confining field of F2 to the lateral lobes and thereby reduce (increase) the confining capacity of the middle (lateral) lobe. Therefore, this type of disturbance often lead to failed eruptions of F1 and F3, while the eruption of F2 should be a successful one. The second type is that the disturbance acts on F1 or F2, which will not lead to any reconnection, and therefore no filament eruption occurs. Here, we present another type of filament eruption in a solar breakout event, in which successive partial and full filament eruptions are involved.

\begin{figure}[b]
\begin{center}
\includegraphics[width=0.8\textwidth]{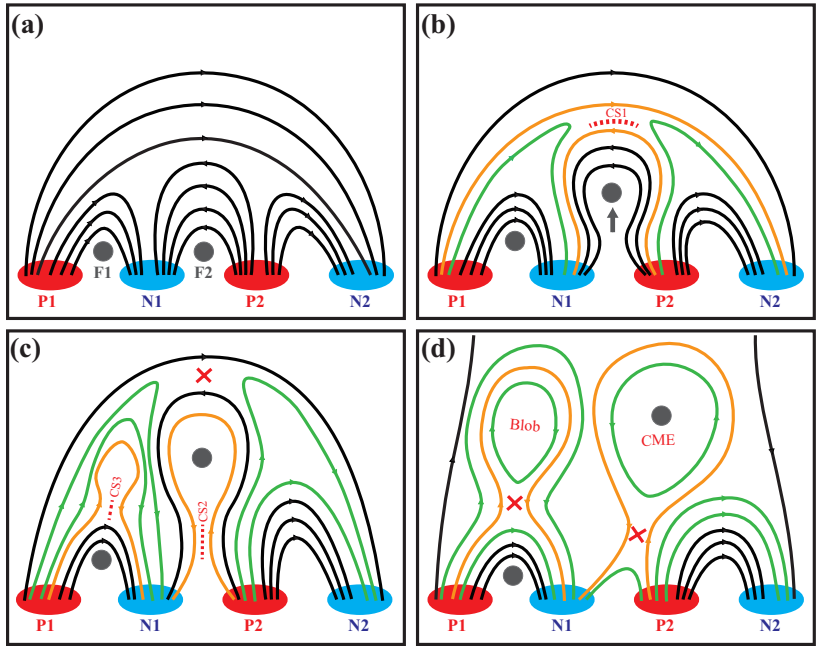} 
\caption{Schematic demonstrating the successive filament eruptions. (a) The initial magnetic configuration. (b) The rising of F2, formation of CS1 and the external reconnection. (c) Formation of CS2 (CS3) underneath (above) F2 (CS3). (d) Reconnections in CS2 and CS3, and the production of the nearly simultaneous CMEs. The arrow pointing to F2 represent the disturbance. The red dotted lines indicate locations of the current sheets, while the reconnection sites are labeled by red ``X'' symbols. The yellow lines represent the field lines to be reconnected, while the green lines are reconnected ones.}
\label{fig2}
\end{center}
\end{figure}

The detailed analysis of the event can be found in \cite{shen12}. Here we just give a brief summery of the results. As shown in Fig.\,\ref{fig2}(a), we find that the extrapolated coronal field above the magnetic source region is of the topology of the breakout model, and the two filaments are located below the middle and the left lobes respectively. The initiation of the successive filament eruptions started from a small mass ejection, which directly interacted with the southern part of F2 and thus resulted in the slow rise of this filament. The slow rise of F2 lasted for about 23 minutes and a speed of 8 km/s. During this period, some signatures for breakout-like external reconnection were observed. For example, two brightening patches at both sides of F1, the appearance of bright loops and a weak hard X-ray source above F1. After the slow rising phase, F2 was quickly accelerated to 102 km/s, and finally, it erupted successfully and caused a CME. The activation of F1 started around the end of F2's slow rising phase, which erupted with strong writhing  motions. When F1 reached its maximum height, the eruption of a blob-like structure was observed above the filament. In the meantime, F1 began to fall back to the solar surface. These results indicate that the eruption of F1 should be a typical partial flux rope eruption. According to the model proposed by \cite{gilb01}, the reconnection site should be located above the filament.

We interpret the observations using the breakout model as shown in Fig.\,\ref{fig2}. Panel (a) presents basic magnetic topology. Due to the disturbance introduced by a small plasma ejection, F2 slowly rises, expanding the middle lobe, which will result in the external magnetic reconnection within the current sheet formed around the coronal null point (see CS1 in panel (b)). According to the magnetic implosion mechanism proposed by \cite{huds00},  the magnetic pressure around the reconnection site will decrease due to energy released during the energy conversion process in coronal transients such as flares. The reduction of magnetic pressure will lead to the contraction of the overlying loop system and the expansion of the low-lying lobes. In addition, the strong writhing of F1 indicates that the eruption of this filament was driven by the kink instability. The reduction of the magnetic tension force of the left lobe facilitates triggering the kink instability within F1. Hence, the magnetic implosion could be a possible physical linkage between the successive filament eruptions within the framework of the breakout model. As the rising of F1 and F2, new current sheet CS2 (CS3) will form underneath (above) F2 (F1). The reconnection within CS2 (CS3) will lead to the successful (partial) eruption of F2 (F1), and the CME (blob) (see panels (c) and (d)). In this model, we can expect two simultaneous CMEs.

\section{Summary}
Based on multi-wavelengths observations, we propose an interpretation for the successive eruptions of two filaments in a solar breakout event. We first introduce the magnetic implosion mechanism to be the physical linkage of the successive filament eruptions. The observations of both the pre-eruptive signatures and the extrapolated three-dimensional coronal fields are in good agreement with the breakout model. Our scenario presented in this article implies the occurrence of nearly simultaneous CMEs. Therefore, this interpretation is important for the forecast of space weather. It should be noted that the breakout scenario is a possible explanation for the observations. We do not intend to exclude other possibilities. In any case, the structural properties of coronal fields are important for producing successive filament eruptions.\\

\noindent {\bf Acknowledgements} This work is supported by the Western Light Youth Project of Chinese Academy of Sciences (CAS), the CAS open research programs (KLSA201204, DMS2012KT008). Y. Shen thank the financial support for young researches to participate the 300th symposium (IAUS300: Nature of Prominences and their role in Space Weather) of the International Astronomical Union.

\end{document}